\documentclass[doublecol]{epl2} 
\usepackage{bbm}
\usepackage{amsmath}
\usepackage{amssymb}
\usepackage{amsthm}

\title{Proposal for an ac spin current source}
\author{Patrick P. Hofer \thanks{E-mail: \email{patrick.hofer@unige.ch}}\inst{1}
 \and Hugo Aramberri \inst{2}
 \and Christoph Schenke \inst{1}
 \and Pierre A. L. Delplace \inst{1}
 }
\shortauthor{Patrick P. Hofer \etal}

\institute{                    
  \inst{1} 
  D\'epartement de Physique Th\'eorique, Universit\'e de Gen\`eve - CH-1211 Gen\`eve 4, Switzerland\\
  \inst{2} Instituto de Ciencia de Materiales de Madrid (ICMM-CSIC) - Cantoblanco 28049 Madrid, Spain
}

\abstract{
We propose an ac current source that can be tuned from a pure {\it charge} to a pure {\it spin} current source. The device consists of two mesoscopic capacitors attached to a two-dimensional strip of a topological insulator. The change from charge to spin current is controlled by an offset in the top gate potentials that drive the capacitors. In addition to this setup, which anticipates the experimental realisation of quantum point contacts in topological insulators, we propose an analogous source in the quantum Hall regime which only relies on presently available building blocks. To this end, we calculate the band structure of a topological insulator in a magnetic field. The intrinsic spin-orbit coupling, together with a split gate, allows for an analogous source, where charge and spin current can be manipulated. The realisation of the device as well as the detection of the ac spin current are within reach of present experimental technology.
}

\begin{document}

\maketitle

\section{Introduction}

Topological insulators (TI) are states of matter which are not characterised by a local order parameter but by a topological invariant that represents a global order of the system
\cite{bernevig:book,kane:2005,hasan:qi}. At the interface between materials featuring different topological invariants, localised states appear. In particular,
two-dimensional TIs which are invariant under time reversal (TR) exhibit pairs of helical edge states at interfaces to topologically trivial insulators such as the vacuum. The
members of these pairs are related by TR and thus propagate in opposite directions and have opposite spin. As supported by recent experiments
\cite{konig:roth,brune:2012,du:spanton,suzuki:grabecki}, the topological nature of these states ensures the absence of backscattering
as long as interactions and inelastic processes are negligible. These properties make edge channels in TIs the ideal candidates for electron waveguides in spintronics applications.

The field of spintronics strives for control and manipulation of the spin degrees of freedom in condensed matter systems \cite{zutic:2004}. Spintronics devices can in general be used as logical base units for high performance computation and their main ingredients are spin transistors and spin filters, based on the spin valve effect \cite{julliere:slonczewski}. Such devices have been realised in various nanostructures, e.g., ferromagnetic tunnel junctions, molecules, and carbon nanotubes
\cite{pasupathy:etal}. More recently, proposals which use the spin polarised edge modes of TIs in order to create spin filters, spin transistors, and spin current pumps
\cite{krueckl:etal} have attracted attention. Particularly promising are logic devices which suggest computation schemes with low power 
consumption \cite{datta:etal}.

Another promising route towards energy efficient computation is provided by electronic few particle processes. Using sources that emit single particles on demand 
\cite{moskalets:book,moskalets:2008,keeling:2008,keeling:2006}, pioneering experiments \cite{feve:bocquillon,mahe:parmentier, bocquillon:2012,dubois:2013} proved the feasibility
of creating and manipulating single electron excitations in a controlled manner.

In this Letter, we propose to use two single-particle sources provided by mesoscopic capacitors \cite{moskalets:2008} to emit electrons and holes into the helical edge states of a TI
\cite{hofer}. Thereby an ac current in the GHz range is produced that can be tuned from a pure charge to a pure spin current. Recent experiments reporting on the detection of ac spin currents in the GHz range \cite{hahn:etal} stress the experimental importance of the field of ac spintronics \cite{jiao:kochan}. A device where a dc current can be tuned from a spin-polarised charge current to a pure spin current is discussed in Ref.~\cite{sothmann:2012}.

Because of spin-orbit coupling, spin is usually not a good
quantum number in the considered systems. The role of spin and total angular momentum is clarified below. The single-particle sources rely on quantum point contacts (QPC),
which represent a considerable experimental challenge in TIs. To circumvent this, we additionally propose a setup where the TI is tuned into the quantum Hall (QH) regime by a strong magnetic field. For this setup, all the building blocks are presently available.
In particular, QPCs are provided by top gates which locally change the filling factor. Since edge states are found at interfaces of regions with a different filling factor, they follow equipotential lines in the QH regime. In TIs, gates can not change the topological order thus QPCs have to be created geometrically.

The quantitative analysis of our proposal is done using material parameters of experimentally available TIs \cite{du:spanton}. Due to the strong intrinsic
spin-orbit coupling that mediates the band inversion in these TIs, the device still acts as a source of pure spin current although TR symmetry is explicitly broken.

The rest of the Letter is structured as follows. First we discuss the edge states in the utilised model, focusing on their spin and angular momentum polarisation. The next section is dedicated to the proposed device in the quantum spin Hall regime. We then present the behaviour of the band structure in a magnetic field, providing the necessary background for the following discussion of the setup in the QH regime. We end the letter with a summary and concluding remarks.

\section{BHZ model}

The model we employ for a two-dimensional TI was developed by Bernevig, Hughes and Zhang (BHZ) \cite{bernevig:2006}. The resulting Bloch-Hamiltonian reads

\begin{equation}
\begin{aligned}
\label{eq:bhz}
&\hspace{1cm}H(\boldsymbol{k})=\begin{pmatrix}
h(\boldsymbol{k}) & 0 \\
0 & h^*(-\boldsymbol{k})
\end{pmatrix},\\
h(\boldsymbol{k})&=\epsilon(\boldsymbol{k})\mathbbm{1}_2+\boldsymbol{d}(\boldsymbol{k})\cdot\boldsymbol{\sigma},\hspace{.85cm}\epsilon(\boldsymbol{k})=\tilde{C}-\tilde{D}k^2\\
\boldsymbol{d}(\boldsymbol{k})&=[\tilde{A}k_x, -\tilde{A}k_y, M(\boldsymbol{k})],\hspace{.5cm}M(\boldsymbol{k})=\tilde{M}-\tilde{B}k^2,
\end{aligned}
\end{equation}
where the quantities with a tilde are material specific parameters, $\mathbbm{1}_2$ is the identity matrix of order $2$, $\boldsymbol{\sigma}$ is the vector of Pauli matrices and
$k^2=k_x^2+k_y^2$. Systems that are described by this Hamiltonian include quantum wells of HgTe/CdTe \cite{bernevig:2006}, InAs/GaSb \cite{liu:2008} and Ge/GaAs \cite{zhang:2013}. 
The basis states of the BHZ Hamiltonian are denoted by $\left(\left|E,+1/2\right>,\left|H,+3/2\right>,\left|E,-1/2\right>,\left|H,-3/2\right>\right)$, where $E$ ($H$) denotes states derived from electron-like (hole-like) bands and the number gives $m_j$, the projection of the total angular momentum on the $z$-axis. These basis states are discussed in detail in Refs.~\cite{lunde:2013,pfeuffer:2000}. Importantly, $m_j$ is a good quantum number for the basis states (and thus for the BHZ Hamiltonian at $k=0$) which does not depend on the specific material or sample geometry. We therefore discuss the proposed device as a source of total angular momentum current. The projection of the electron spin onto the $z$-axis, $m_s$, is only a good quantum number for the states $\left|H,\pm3/2\right>$ which have $m_s=\pm1/2$. However, in the case of HgTe/CdTe, one can make an estimate of the spin polarisation of the $m_j=\pm1/2$ states based on Ref.~\cite{pfeuffer:2000}, which gives $\left<E,\pm1/2\right|S_z\left|E,\pm1/2\right>\approx\pm0.73\hbar/2$ for a well-width of $7\:$nm. Here, $S_z$ is the $z$-component of the electron-spin operator. For all basis states, we thus find that $\langle S_z\rangle$ and $m_j$ have equal signs, which means that the currents associated with these two quantities are proportional to each other. This justifies the notion of spin current although we discuss the source in terms of total angular momentum current which is sample and material independent unlike the electron-spin current.

We refer to the eigenstates of the upper (lower) block of Eq.~\eqref{eq:bhz} as spin up (down) states. The edge modes can be found by solving the Hamiltonian for hard wall boundary conditions \cite{konig:2008,zhou:2008}. Since they have equal weights on the $\left|E,\pm1/2\right>$ and $\left|H,\pm3/2\right>$ state, we find $\left<J_z\right>_\alpha=\pm\hbar$, where $J_z$ is the $z$-component of the total angular momentum operator and $\alpha$ labels the spin of the edge state. Tuning the Fermi energy, $E_F$, into the bulk gap, only these edge modes contribute to transport. They constitute the scattering channels which are referred to below.

\section{Time reversal invariant source}

\begin{figure}[t!]
\onefigure[width=.9\columnwidth]{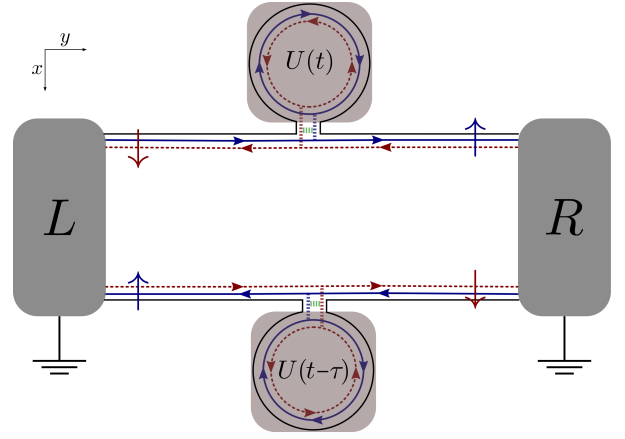}
\caption{Proposed ac spin current source in the topological insulator regime. Two mesoscopic capacitors are attached to a two-dimensional topological insulator. If they are operated
in phase ($\tau=0$), they both emit electrons and holes at equal times resulting in a pure and equal ac {\it charge} current in the two contacts ($R/L$). A delay, $\tau$, can synchronise electron and hole emission on the different edges. This results in a pure ac {\it spin} current of opposite sign in the different contacts. The arrows indicate the expectation value of the spin.}
\label{fig:source_ti}
\end{figure}

The proposed source is sketched in Fig.~\ref{fig:source_ti}. It consists of two mesoscopic capacitors attached to opposite sides of a strip of a two-dimensional TI. The working principle of a mesoscopic capacitor in the TI regime is discussed in Refs.~\cite{hofer} and is similar to the analogous device in the QH regime \cite{moskalets:2008}, which has been implemented experimentally \cite{feve:bocquillon}. A mesoscopic capacitor consists of a quantum dot, tunnel coupled to the helical edge states. Using a top gate, $U(t)$, a pair of Kramers degenerate quantum dot levels are moved above and below $E_F$ in a
periodic manner. Every time the Kramers pair is lifted above $E_F$, one electron is ejected into each outgoing channel, one propagating to the left, one to the right. Every time the
level pair moves below $E_F$ a hole is emitted into each channel. A single capacitor thereby creates an equal, spin-polarised ac charge current in the contacts $L$ and $R$.

For simplicity, identical capacitors are considered, the only difference being a time shift $\tau$ in their periodic top gate potential. If the two capacitors are operated in phase ($\tau=0$), the charge current measured at one of the contacts ($R/L$) is doubled with respect to a single capacitor but the spin current is cancelled, since there is a channel of each spin entering each contact. Operating the capacitors out of phase, such that electron emission of one capacitor is synchronised with hole emission of the other capacitor, the charge current vanishes. Since an electron of a given spin species and a hole of the opposite spin species contribute equally to the spin current, this results in a pure ac spin current which is opposite in each contact.

For a quantitative analysis, we resort to the scattering approach \cite{moskalets:book}. There are four channels connecting the two contacts in the setup sketched in Fig.~\ref{fig:source_ti}. Since the channels at the upper and lower edges are well separated and because of TR invariance, there is no scattering between the channels. The total charge current is therefore just the sum of the currents created by each capacitor. In order to move a dot level above and below $E_F$, we consider the experimentally relevant step-like top gate potential with an amplitude given by the dot level spacing $\Delta$ and a frequency $2\pi/\mathcal{T}$ in the GHz range \cite{mahe:parmentier}. The charge current emitted into a single channel by the capacitor located at the upper edge is periodic in time with period $\mathcal{T}$ and reads \cite{keeling:2008}

\begin{equation}
\label{eq:curr}
I^c(t)=\begin{cases}\frac{e}{\tau_D}e^{-t/\tau_D}\hspace{1.42cm}\text{for}\hspace{.2cm}0\leq t<\mathcal{T}/2, \\
-\frac{e}{\tau_D}e^{-(t-\mathcal{T}/2)/\tau_D}\hspace{.2cm}\text{for}\hspace{.2cm}\mathcal{T}/2\leq t<\mathcal{T}.
\end{cases}
\end{equation}
Here $\tau_D=h/(D\Delta)$ is the dwell time with $D$ being the transmission probability through the QPC and $e<0$ is the elementary charge. We consider the case where $\tau_D\ll\mathcal{T}/2$, such that in each cycle exactly one electron and one hole is emitted. The charge current emitted by the capacitor located at the lower edge is given by $I^c(t-\tau)$.

Due to the linear dispersion, the spin current in a given channel $\alpha$ is proportional to the charge current $I^s_\alpha(t)=\left<J_z\right>_\alpha I^c_\alpha/e$. The quantised emission of a single particle into channel $\alpha$ is therefore accompanied by the quantised emission of angular momentum in units of $\left<J_z\right>_\alpha$.
Operating the two capacitors with a time shift $\tau$ leads to the currents in the right and left contacts

\begin{equation}
\label{eq:currents}
\begin{aligned}
&I^c_R(t)=I^c_L(t)=I^c(t)+I^c(t-\tau),\\
&I^s_R(t)=-I^s_L(t)=\frac{\hbar}{e}\left[I^c(t)-I^c(t-\tau)\right],
\end{aligned}
\end{equation}
where the subscript $R/L$ denotes the contact and we made use of the fact that $\left<J_z\right>_\alpha=\pm\hbar$ for the edge channels.

\begin{figure}[t]
\onefigure[width=.9\columnwidth]{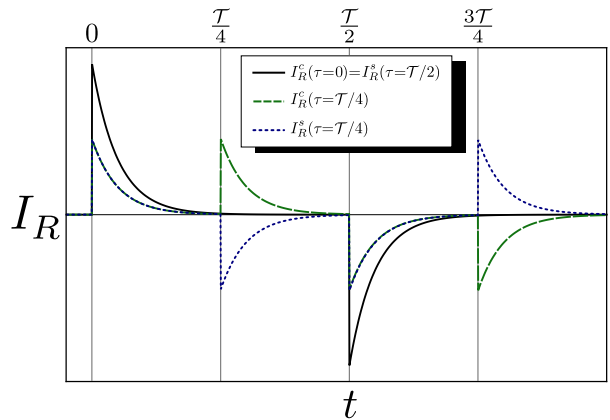}
\caption{Charge and spin current ($e=\hbar=1$) emitted by the source into the right contact. Black (solid) line shows the pure charge current for synchronous operation which is equal to the pure
spin current for $\tau=\mathcal{T}/2$. The green (broken) line shows the charge current, the blue (dashed) line the spin current, for a finite delay. Changing $\tau$ moves the
peaks and dips centred around $\tau$ and $\mathcal{T}/2+\tau$, going from a pure charge ($\tau=0$) to a pure spin current ($\tau=\mathcal{T}/2$). Note that an equal (opposite) charge (spin)
current enters the left contact.}
\label{fig:spincharge}
\end{figure}

The charge and spin currents entering the right contact are plotted in Fig.~\ref{fig:spincharge}. For $\tau=0$, each capacitor emits an electron at time $t=0$ and a hole at time $t=\mathcal{T}/2$. The charge current is thus maximised and there is no spin current. For a finite $\tau$, the upper capacitor emits an electron at time $t=0$ which contributes positively to both charge and spin current, while the lower capacitor emits an electron at time $t=\tau$ which contributes positively to the charge and negatively to the spin current. The opposite holds for the holes and for the spin current in the left contact. Finally, when $\tau=\mathcal{T}/2$, the electron of the upper capacitor is synchronised with the hole of the lower capacitor. The charge current vanishes and the spin current is maximised. Controlling the time shift thus allows one to tune from a pure charge to a pure spin current. Conveniently, mesoscopic capacitors are usually operated in the GHz range, where experimental detection of ac spin currents has recently been reported \cite{hahn:etal}.

\section{Edge states in a magnetic field}

\begin{figure*}[t!]
\onefigure[width=\textwidth]{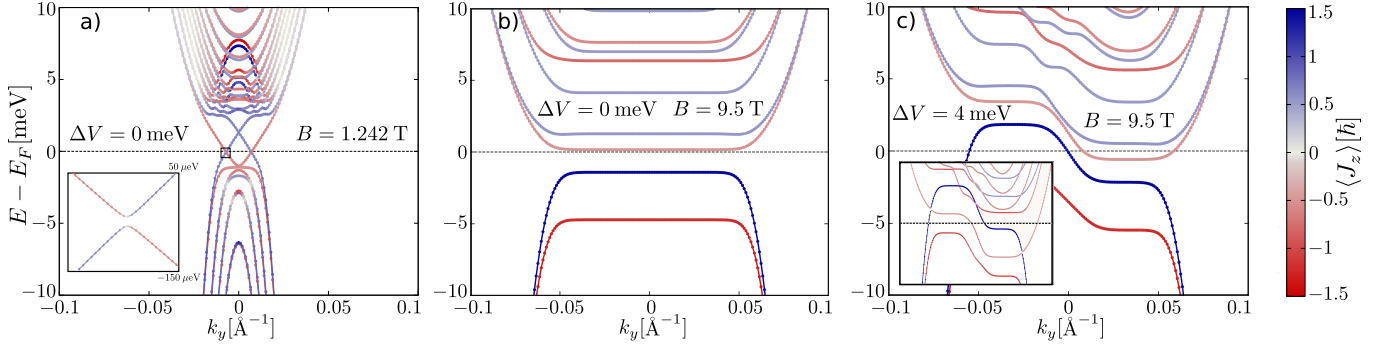}
\caption{Evolution of the band structure of InAs/GaSb in a magnetic field. The colour scheme indicates the expectation value $\left<J_z\right>$. a) For a weak magnetic field, spin up (down) states move up (down) in energy and the crossing of opposite spin states moves to negative (positive) $k_y$ for the upper (lower) edge. The
crossings at zero energy are lifted by the inversion asymmetry terms as shown in the inset. b) In a strong magnetic field, Landau levels develop with a characteristic texture of $\left<J_z\right>$. For fields stronger then $B_c$, the intrinsic band inversion is revoked and all the electron-like Landau levels are above the hole-like Landau levels. c) A 
split gate can move the band structure up (down) on the upper (lower) half of the sample by locally applying the potential $V_u$ ($V_l$), with $\Delta V\equiv V_u-V_l$. The upper half then supports a $\left<J_z\right>_u=3\hbar/2$ and the lower half a $\left<J_z\right>_l=-\hbar/2$ edge channel. The inset shows the band structure including the Zeeman splitting for the same magnetic field and $\Delta V=7$\,meV.} For these calculations we used a tight-binding regularization of the BHZ model.
\label{fig:bfield}
\end{figure*}

Since the fabrication of QPCs in TIs remains an open challenge, we extend our proposal to the QH regime, where QPCs can be implemented using gates. To this end, we first describe the behaviour of the helical edge states in a perpendicular magnetic field. The transition from the quantum spin Hall to the QH regime in the honeycomb lattice is discussed in Refs.~\cite{shevtsov:2012}, the effect of a magnetic field in HgTe/CdTe quantum wells in Refs.~\cite{scharf:chen}. In the light of recent experiments \cite{du:spanton}, we consider an InAs/GaSb quantum well with the material parameters given in Ref.~\cite{wang:2013}. The band structure evolution in a magnetic field is qualitatively similar for all materials described by the BHZ model. However, the required magnetic field depends on the material parameters. It is quantitatively similar in HgTe/CdTe while in Ge/GaAs it is two orders of magnitude higher. Since we are interested in the spin texture, we take into account the spin coupling terms that arise in this material due to bulk (BIA) and structural inversion asymmetry (SIA) \cite{liu:2008}

\begin{equation}
\label{eq:bia}
H_{\rm BIA}(\boldsymbol{k})=\begin{pmatrix}
0 & 0 & \Delta_ek_+ & -\Delta\\
0 & 0 & \Delta & \Delta_hk_-\\
\Delta_ek_- & \Delta & 0 & 0\\
-\Delta & \Delta_hk_+ & 0 & 0
\end{pmatrix},
\end{equation}
and

\begin{equation}
\label{eq:sia}
H_{\rm SIA}(\boldsymbol{k})=\begin{pmatrix}
0 & 0 & iRk_- & 0\\
0 & 0 & 0 & 0\\
-iR^*k_+ &0 & 0 & 0\\
0 & 0 & 0 & 0
\end{pmatrix},
\end{equation}
where $k_\pm=k_x\pm ik_y$ and $\Delta_{e/h}$, $R$ are material dependent constants. We verified numerically that the Zeeman splitting only quantitatively modifies those features required for the generation of spin currents (see the inset in Fig.~\ref{fig:bfield}\,c) and thus neglect this effect in the general discussion, returning to it at the end of this letter. The Hamiltonian we consider is thus the sum of Eqs.~(\ref{eq:bhz}, \ref{eq:bia}, \ref{eq:sia}), where the magnetic field is introduced using the minimal coupling in the Landau gauge $k_x\rightarrow -i\partial_x$, $k_y\rightarrow k_y-eBx/\hbar$.

The behaviour of the edge states in a magnetic field is illustrated in Fig.~\ref{fig:bfield}\,a and b and in the supplemental video on-line.
For small magnetic fields, the up (down) spin states are pushed up (down) in energy. This shifts the crossing of the opposite spin states at zero energy to negative $k_y$ values for the upper edge and to positive $k_y$ values for the lower edge. Since these crossings are no longer protected by TR invariance and since the spins are coupled by the inversion asymmetry terms, a gap opens up (see inset of Fig.~\ref{fig:bfield}\,a).

Further increasing the magnetic field moves the crossings between the same spin states into the bulk states. Coupling of the different edges via the bulk states lifts these degeneracies and Landau levels are formed. The spin up states give rise to a hole-like Landau level which is above the electron-like spin-down Landau level due to the intrinsic band inversion (not shown). Increasing the magnetic field further shifts the hole-like spin-up level down and the electron-like spin-down level up in energy, until they eventually undergo an avoided crossing at $B_c=\hbar\tilde{M}/(e\tilde{B})\approx8.23$\:T \cite{konig:2008}. For higher magnetic fields, all electron-like Landau levels are above the hole-like Landau levels and they alternate in the sign of $\left<J_z\right>_\alpha$ (see Fig.~\ref{fig:bfield}\,b). Since at high magnetic fields, $k_y$ is directly proportional to the $x$-component of the centre of cyclotron motion, we can read off the localisation of the electrons directly from the band structure.

\section{Quantum Hall source}
\begin{figure}[t]
\onefigure[width=.9\columnwidth]{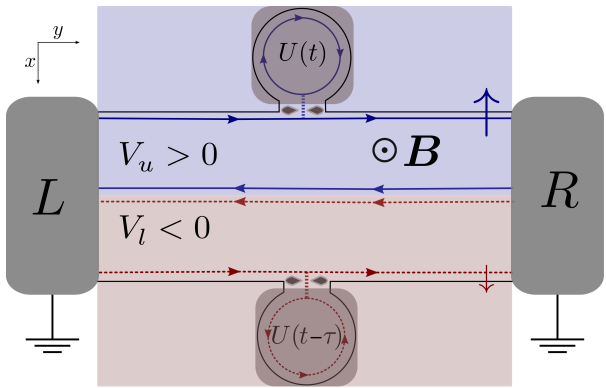}
\caption{Proposed ac spin current source in the quantum Hall regime. A strong magnetic field is used to produce Landau levels. Using a split gate ($V_u$ and $V_l$, blue and red shading), the upper half
is tuned into a regime where one edge state with $\left<J_z\right>_u=3\hbar/2$ (blue, solid) propagates clockwise, while in the lower half an edge state with
$\left<J_z\right>_l=-\hbar/2$ (red, dashed) propagates counter clockwise. The leftmoving states are localised in the middle of the sample, ensuring the absence of backscattering.}
\label{fig:source_exp}
\end{figure}

In this section, we extend our proposal to the QH regime, where it consists exclusively of experimentally available building blocks.

A bottom gate can be used to shift the band structure relative to $E_F$. As experimentally demonstrated in Ref.~\cite{brune:2012}, a split gate can be used to shift the band structure up in energy in one half, and down in energy in the other half of the sample. We model this by including a potential $V$ which is positive in the upper half and negative in the lower half of the sample (see Fig.~\ref{fig:source_exp}). This leads to the band structure shown in Fig.~\ref{fig:bfield}\,c with four states at $E_F$. The states at the edges of the sample both propagate to the right and have spin polarisations of opposite signs but different magnitudes. The state at the upper edge has $\left<J_z\right>_u=3\hbar/2$, the state at the lower edge $\left<J_z\right>_l=-\hbar/2$. Since the leftmoving states are located in the middle of the sample, there is no backscattering.

To achieve a source of ac spin current, a mesoscopic capacitor is attached to each side of the sample as sketched in Fig.~\ref{fig:source_exp}. The current emitted by the capacitors is described by Eq.~\eqref{eq:curr}. Note that in this setup, the channels connected to the mesoscopic capacitors both propagate to the right, meaning that $I_L^c=I_L^s=0$. The charge current in the right contact is still given by Eq.~\eqref{eq:currents} and the spin current reads 

\begin{equation}
\label{eq:currentspin}
I^s_R(t)=\frac{\hbar}{e}\left[\frac{3}{2}I^c(t)-\frac{1}{2}I^c(t-\tau)\right].
\end{equation}
Even at $\tau=0$, this source thus emits a spin current of $I^c$; a pure charge current can no longer be created. However, a pure spin current of $2I^c$ can still be generated by an offset $\tau=\mathcal{T}/2$.

We end this letter with some remarks on the feasibility of our proposal.
As shown in Fig.~\ref{fig:bfield}\,c, the spin-down state that is coupled to the lower capacitor has a spin-up state in its vicinity. The gap between these states is approximately $1\:$meV. Even if the Fermi energy lies above the spin-up Landau level, the QPC of the capacitor can be operated to be transmitting only for the outer, spin-down edge channel.

In the inset of Fig.~\ref{fig:bfield}\,c, we include the Zeeman splitting with estimates for the $g$-factor of $-8$ for electrons and $-3$ for holes \cite{nilsson:2006}. The Zeeman splitting has no observable effect on the spin polarisation. Although the band inversion is no longer revoked at $9.5$\,T, the available states at $E_F$, and thus our results, remain unchanged.
In the contacts, a magnetic field as high as $9.5$\:T will induce a Zeeman splitting which even for high $g$-factors remains much smaller than $E_F$, justifying the approximation of a linear spectrum. Due to this splitting, spin-flip processes need to be inelastic or accompanied by a momentum transfer, making them less likely.

Recent work indicates that our proposal is robust against disorder and deformations of the quantum dots \cite{xing:2014}.
Although we focused on InAs/GaSb quantum wells, our results are quantitatively similar for HgTe/CdTe quantum wells making our proposal relevant for both experimentally available TI materials \cite{konig:roth,du:spanton}.

\section{Conclusion}

Using two mesoscopic capacitors \cite{moskalets:2008} attached to the sides of a TI, we propose an ac current source that can be tuned from a pure {\it charge} to a pure {\it spin} current source by varying an offset in the driving potentials of the capacitors. So far, QPCs have not been implemented in TIs. We therefore extend our proposal to the QH regime, where QPCs are provided by gates. To this end, we discuss the behaviour of the helical edge states in a perpendicular magnetic field \cite{scharf:chen}. The emerging Landau levels have a characteristic angular momentum structure due to the intrinsic spin-orbit coupling of TI materials. Using a split gate \cite{brune:2012}, we produce a situation where two co-propagating states with angular momentum expectation value of opposite directions are localised at the opposite edges of the sample. Analogous to the source in the TR invariant regime, the charge and spin currents can be manipulated by varying the offset in the driving potentials. Since the magnitude of the angular momentum expectation value on the two edges is different, a pure {\it charge} current cannot be produced any more. However, a pure ac {\it spin} current can still be generated.

Due to recent experimental breakthroughs in the fields of single electron sources \cite{feve:bocquillon}, topological insulators \cite{konig:roth,du:spanton}, and spin current detection \cite{hahn:etal}, our proposal is experimentally feasible and will hopefully stimulate more experiments in the upcoming field of ac spintronics.

%\begin{table}
%\caption{Table caption.}
%\label{tab.1}
%\begin{center}
%\begin{tabular}{lcr}
%first  & table & row\\
%second & table & row
%\end{tabular}
%\end{center}
%\end{table}

\acknowledgments
We acknowledge the involvement of M. B\"uttiker at an early stage. We thank G. F\`eve, J. Li, M. Moskalets and the referee for valuable comments. All numerical calculations were done using the MOJITO code written by J. Li.

This work is financially supported by the SNSF. H.A. acknowledges support from the Spanish Ministerio de Ciencia e Innovacion (MICINN) through the Grant No. MAT2009-14578-C03-03.


\begin{thebibliography}{10}
\expandafter\ifx\csname url\endcsname\relax\def\url#1{\texttt{#1}}\fi

\bibitem{bernevig:book}
\Name{Bernevig B.~A. \and Hughes T.~L.} \Book{Topological Insulators and
  Topological Superconductors} (Princeton University Press) 2013.

\bibitem{kane:2005}
\Name{Kane C.~L. \and Mele E.~J.} \REVIEW{Phys. Rev. Lett.}{95}{2005}{146802}.

\bibitem{hasan:qi}
\Name{Hasan M.~Z. \and Kane C.~L.} \REVIEW{Rev. Mod. Phys.}{82}{2010}{3045};
\Name{Qi X.-L. \and Zhang S.-C.} \REVIEW{ibid.}{83}{2011}{1057}.

\bibitem{konig:roth}
\Name{K\"onig M., Wiedmann S., Br\"une C., Roth A., Buhmann H., Molenkamp
  L.~W., Qi X.-L. \and Zhang S.-C.} \REVIEW{Science}{318}{2007}{766};
\Name{Roth A., Br\"une C., Buhmann H., Molenkamp L.~W., Maciejko J., Qi X.-L.
  \and Zhang S.-C.} \REVIEW{ibid.}{325}{2009}{294}.

\bibitem{brune:2012}
\Name{Br\"une C., Roth A., Buhmann H., Hankiewicz E.~M., Molenkamp L.~W.,
  Maciejko J., Qi X.-L. \and Zhang S.-C.} \REVIEW{Nat Phys}{8}{2012}{485}.

\bibitem{suzuki:grabecki}
\Name{Suzuki K., Harada Y., Onomitsu K. \and Muraki K.} \REVIEW{Phys. Rev.
  B}{87}{2013}{235311};
\Name{Grabecki G., Wr\'obel J., Czapkiewicz M., Cywi\ifmmode~\acute{n}\else
  \'{n}\fi{}ski L., Giera\l{}towska S., Guziewicz E., Zholudev M., Gavrilenko
  V., Mikhailov N.~N., Dvoretski S.~A., Teppe F., Knap W. \and Dietl T.}
  \REVIEW{ibid.}{88}{2013}{165309}.

\bibitem{du:spanton}
\Name{Du L., Knez I., Sullivan G. \and Du R.-R.} arXiv:{1306.1925}
 (2013);
\Name{Spanton E.~M., Nowack K.~C., Du L., Du R.-R. \and Moler K.~A.}
  arXiv:{1401.1531}
 (2014).

\bibitem{zutic:2004}
\Name{\ifmmode \check{Z}\else \v{Z}\fi{}uti\ifmmode~\acute{c}\else \'{c}\fi{}
  I., Fabian J. \and Das~Sarma S.} \REVIEW{Rev. Mod. Phys.}{76}{2004}{323}.

\bibitem{julliere:slonczewski}
\Name{Julliere M.} \REVIEW{Phys. Lett. A}{54}{1975}{225 };
\Name{Slonczewski J.~C.} \REVIEW{Phys. Rev. B}{39}{1989}{6995}.

\bibitem{pasupathy:etal}
\Name{Pasupathy A.~N., Bialczak R.~C., Martinek J., Grose J.~E., Donev L.
  A.~K., McEuen P.~L. \and Ralph D.~C.} \REVIEW{Science}{306}{2004}{86};
  \Name{Cottet A., Kontos T., Sahoo S., Man H.~T., Choi M.-S., Belzig W., Bruder
    C., Morpurgo A.~F. \and Sch\"onenberger C.} \REVIEW{Semicond. Sci.
    Technol.}{21}{2006}{S78};
    \Name{Schelp L.~F., Fert A., Fettar F., Holody P., Lee S.~F., Maurice J.~L.,
      Petroff F. \and Vaur\`es A.} \REVIEW{Phys. Rev. B}{56}{1997}{R5747};
\Name{Man H.~T., Wever I. J.~W. \and Morpurgo A.~F.} \REVIEW{ibid.}{73}{2006}{241401}.

\bibitem{krueckl:etal}
\Name{Krueckl V. \and Richter K.} \REVIEW{Phys. Rev. Lett.}{107}{2011}{086803};
\Name{Maciejko J., Kim E.-A. \and Qi X.-L.} \REVIEW{Phys. Rev.
  B}{82}{2010}{195409};
\Name{Mahfouzi F., Nikoli\ifmmode~\acute{c}\else \'{c}\fi{} B.~K., Chen S.-H.
 \and Chang C.-R.} \REVIEW{ibid.}{82}{2010}{195440};
\Name{Dolcetto G., Cavaliere F., Ferraro D. \and Sassetti M.} \REVIEW{ibid.}{87}{2013}{085425};
\Name{Ferraro D., Dolcetto G., Citro R., Romeo F.  \and Sassetti M.} \REVIEW{ibid.}{87}{2013}{245419}.

\bibitem{datta:etal}
\Name{Datta S. \and Das B.} \REVIEW{Appl. Phys. Lett.}{56}{1990}{665};
\Name{Wolf S.~A., Awschalom D.~D., Buhrman R.~A., Daughton J.~M., von Molnár
  S., Roukes M.~L., Chtchelkanova A.~Y. \and Treger D.~M.}
  \REVIEW{Science}{294}{2001}{1488};
\Name{Behin-Aein B., Datta D., Salahuddin S. \and Datta S.} \REVIEW{Nat.
  Nanotechnol.}{5}{2010}{266}.

\bibitem{moskalets:book}
\Name{Moskalets M.~V.} \Book{Scattering matrix approach to non-stationary
  quantum transport} (Imperial {C}ollege {P}ress) 2012.

\bibitem{moskalets:2008}
\Name{Moskalets M., Samuelsson P. \and B\"uttiker M.} \REVIEW{Phys. Rev.
  Lett.}{100}{2008}{086601}.

\bibitem{keeling:2008}
\Name{Keeling J., Shytov A.~V. \and Levitov L.~S.} \REVIEW{Phys. Rev.
  Lett.}{101}{2008}{196404}.

\bibitem{keeling:2006}
\Name{Keeling J., Klich I. \and Levitov L.~S.} \REVIEW{Phys. Rev.
  Lett.}{97}{2006}{116403}.

\bibitem{feve:bocquillon}
\Name{F\`eve G., Mah\'e A., Berroir J.-M., Kontos T.,
  Pla\ifmmode~\mbox{\c{c}}\else \c{c}\fi{}ais B., Glattli D.~C., Cavanna A.,
  Etienne B. \and Jin Y.} \REVIEW{Science}{316}{2007}{1169};
\Name{Bocquillon E., Freulon V., Berroir J.-M., Degiovanni P., Pla\c{c}ais B.,
  Cavanna A., Jin Y. \and F\`eve G.} \REVIEW{ibid.}{339}{2013}{1054}.

\bibitem{mahe:parmentier}
\Name{Mah\'e A., Parmentier F.~D., Bocquillon E., Berroir J.-M., Glattli D.~C.,
  Kontos T., Pla\ifmmode~\mbox{\c{c}}\else \c{c}\fi{}ais B., F\`eve G., Cavanna
  A. \and Jin Y.} \REVIEW{Phys. Rev. B}{82}{2010}{201309};
  \Name{Parmentier F.~D., Bocquillon E., Berroir J.-M., Glattli D.~C.,
    Pla\ifmmode~\mbox{\c{c}}\else \c{c}\fi{}ais B., F\`eve G., Albert M., Flindt
    C. \and B\"uttiker M.} \REVIEW{ibid.}{85}{2012}{165438}.

\bibitem{bocquillon:2012}
\Name{Bocquillon E., Parmentier F.~D., Grenier C., Berroir J.-M., Degiovanni
  P., Glattli D.~C., Pla\ifmmode~\mbox{\c{c}}\else \c{c}\fi{}ais B., Cavanna
  A., Jin Y. \and F\`eve G.} \REVIEW{Phys. Rev. Lett.}{108}{2012}{196803}.


\bibitem{dubois:2013}
\Name{Dubois J., Jullien T., Portier F., Roche P., Cavanna A., Jin Y.,
  Wegscheider W., Roulleau P. \and Glattli D.~C.}
  \REVIEW{Nature}{502}{2013}{659}.

\bibitem{hofer}
\Name{Hofer P.~P. \and B\"uttiker M.} \REVIEW{Phys. Rev.
  B}{88}{2013}{241308(R)};
\Name{Inhofer A. \and Bercioux D.} \REVIEW{ibid.}{88}{2013}{235412}.

\bibitem{hahn:etal}
\Name{Hahn C., de~Loubens G., Viret M., Klein O., Naletov V.~V. \and
  Ben~Youssef J.} \REVIEW{Phys. Rev. Lett.}{111}{2013}{217204};
\Name{Wei D., Obstbaum M., Ribow M., Back C.~H. \and Woltersdorf G.} \REVIEW{Nat. Commun.}{5}{2014}{3768};
\Name{Hyde P., Bai L., Kumar D., Southern B., Huang S.~Y., Miao B.~F., Chien
  C.~L. \and Hu C.-M.}  \REVIEW{Phys. Rev. B}{89}{2014}{180404(R)};
\Name{Weiler M., Shaw J.~M., Nembach H.~T. \and Silva T.~J.} arXiv:{1401.6469}
 (2014).

\bibitem{jiao:kochan}
\Name{Jiao H. \and Bauer G. E.~W.} \REVIEW{Phys. Rev.
  Lett.}{110}{2013}{217602};
\Name{Kochan D., Gmitra M. \and Fabian J.} \REVIEW{ibid.}{107}{2011}{176604}.

\bibitem{sothmann:2012}
\Name{Sothmann B. \and B\"uttiker M.} \REVIEW{EPL (Europhys.
  Lett.)}{99}{2012}{27001}.

\bibitem{bernevig:2006}
\Name{Bernevig B.~A., Hughes T.~L. \and Zhang S.-C.}
  \REVIEW{Science}{314}{2006}{1757}.

\bibitem{liu:2008}
\Name{Liu C., Hughes T.~L., Qi X.-L., Wang K. \and Zhang S.-C.} \REVIEW{Phys.
  Rev. Lett.}{100}{2008}{236601}.

\bibitem{zhang:2013}
\Name{Zhang D., Lou W., M. M., Zhang S.-C. \and Chang K.} \REVIEW{Phys. Rev.
  Lett.}{111}{2013}{156402}.

\bibitem{lunde:2013}
\Name{Lunde A.~M. \and Platero G.} \REVIEW{Phys. Rev. B}{88}{2013}{115411}.

\bibitem{pfeuffer:2000}
\Name{Pfeuffer-Jeschke A.} Ph.D. thesis Physikalisches Institut, Universit\"at
  W\"urzburg (2000).

\bibitem{konig:2008}
\Name{K\"{o}nig M., Buhmann H., Molenkamp L.~W., Hughes T., Liu C.-X., Qi X.-L.
  \and Zhang S.-C.} \REVIEW{J. Phys. Soc. Jpn.}{77}{2008}{031007}.

\bibitem{zhou:2008}
\Name{Zhou B., Lu H.-Z., Chu R.-L., Shen S.-Q. \and Niu Q.} \REVIEW{Phys. Rev.
  Lett.}{101}{2008}{246807}.

\bibitem{shevtsov:2012}
\Name{Shevtsov O., Carmier P., Petitjean C., Groth C., Carpentier D. \and
  Waintal X.} \REVIEW{Phys. Rev. X}{2}{2012}{031004};
  \Name{Beugeling W., Goldman N. \and
    Morais Smith C.} \REVIEW{Phys. Rev. B}{86}{2012}{075118}.

\bibitem{scharf:chen}
\Name{Scharf B., Matos-Abiague A. \and Fabian J.} \REVIEW{Phys. Rev.
  B}{86}{2012}{075418};
\Name{Chen J.-C., Wang J. \and Sun Q.-F.} \REVIEW{ibid.}{85}{2012}{125401}.

\bibitem{wang:2013}
\Name{Wang Q., Liu X., Zhang H.-J., Samarth N., Zhang S.-C. \and Liu C.-X.} arXiv:{1311.4113}
 (2013).
 
\bibitem{nilsson:2006}
\Name{Nilsson K., Zakharova A., Lapushkin I., Yen S. T., \and Chao K. A.} \REVIEW{Phys. Rev.
  B}{74}{2006}{075308}.

\bibitem{xing:2014}
\Name{Xing Y., Sun Q.-F. \and Wang J.} arXiv:{1403.7125}
 (2014).

\end{thebibliography}
\end{document}